\documentclass[sn-mathphys-num]{sn-jnl}
\usepackage{graphicx}%
\usepackage{amsmath,amssymb,amsfonts}%
\usepackage{amsthm}%
\usepackage{mathrsfs}%
\usepackage[title]{appendix}%
\usepackage{xcolor}%
\usepackage{textcomp}%
\usepackage{manyfoot}%
\usepackage{booktabs}%
\usepackage{algorithm}%
\usepackage{algorithmicx}%
\usepackage{algpseudocode}%
\usepackage{listings}%
\usepackage{threeparttable}
\usepackage{amsmath}
\usepackage{graphicx}
\usepackage{dcolumn}
\usepackage{bm}
\usepackage{caption}
\usepackage{float}
\usepackage{dsfont}
\usepackage{mathrsfs}
\RequirePackage{mathptmx} 
\usepackage{xspace}
\usepackage{etoolbox}
\usepackage{caption}
\usepackage{array}
\usepackage{booktabs}
\usepackage{multirow}
\usepackage{makecell}
\theoremstyle{thmstyleone}%

\theoremstyle{thmstyletwo}%
\theoremstyle{thmstylethree}%
\begin{document}
\title[Article Title]{Study of $\beta$-Decay, log $\textit{ft}$ Values and Nuclear Structure Properties of Neutron-rich Ge Nuclei}
\author[1] {\fnm{Jameel-Un} \sur{Nabi}}
\author[1] {\fnm{Wajeeha} \sur{Khalid}} 
\author[2] {\fnm{Abdul} \sur{Kabir}}
\author[1] {\fnm{Syeda Anmol} \sur{Rida}}
\affil[1]{\orgdiv{Department of Physics}, \orgname{University of Wah}, \orgaddress{\street{Quaid Avenue}, \city{Wah Cantt}, \postcode{47040}, \state{Punjab}, \country{Pakistan}}}
\affil[2] {\orgdiv{Department of Space Science}, \orgname{Institute of Space Technology},   \orgaddress{\street{} \city{Islamabad}, \postcode{44000}, \state{Punjab}, \country{Pakistan}}}

\abstract
{We use the relativistic mean field (RMF) model to conduct a thorough analysis of the ground-state properties of  $^{67-80}$Ge nuclei.  Binding energies and  neutron skin thicknesses are computed for a total of 14 neutron-rich  Ge isotopes. This study covers a comprehensive overview of the state of the art regarding the RMF model's explanation of nuclear ground-state properties. Furthermore, we examine the log $\textit{ft}$ values of the allowed $\beta^\pm$ decays and electron/positron capture rates on Ge isotopes in the mass region $A$ = 67--80.
The proton-neutron quasiparticle random phase approximation (pn-QRPA) method was employed to calculate the $\beta$ decay properties of selected Ge isotopes. The calculated log $\textit{ft}$ values show good consistency with the measured data.  The predicted $\beta$ decay half-lives are within a factor of 2 of the measured values. For high temperature and density values prevailing in the core of massive stars, the calculated pn-QRPA  stellar rates are up to an order of magnitude bigger than the independent shell model rates. The findings reported in the current investigation could prove valuable for simulating the late-stage stellar evolution of massive stars.}  




\maketitle
\newpage
\section{Introduction}\label{sec1}

Nuclear processes that take place in  the stellar environment have a significant impact on element synthesis. During  evolution of a star, numerous types of fusion reactions consume  hydrogen, helium, carbon, oxygen, neon and silicon  as ingredients to convert hydrogen to iron group nuclei. The binding energy per nucleon is 8.7 MeV through fusion of nuclei. During stellar evolution, chemical modifications are driven primarily by processes that are hardly able to proceed under existing physical conditions. These alterations take place over periods spanning millions to billions of years. Higher temperatures, which result in higher bombardment energy and bigger cross sections, are experienced by reactions that take place in explosive environments.  Proton captures are permitted in explosive settings, up to the proton drip-line, in a hydrogen-rich environment. This results in the rapid proton-capture process  ($rp$-process). Conversely, neutron captures reaching up to the neutron drip line require highly neutron-rich environments, which are also the locations where the rapid neutron-capture ($r$-process) occurs.~\cite{Arn73}. 
Nuclei heavier than iron are mostly formed by the $r$-process. At the same time, this heavy nucleosynthesis process plays an integral role in the neutronization of massive stars and creation of heavier isotopes. The $r$-process occurs in high temperature ($T\,\approx$\,$10^{9}$\,K) and high neutron density ($\rho_n > 10^{20}$\,g\,cm$^{-3}$) environments~\cite{Bur57, Woo94}.  A better understanding of the $r$-process requires a reliable calculation of the  $\beta$  decay of neutron-rich nuclei under stellar conditions.
For neutron-rich nuclei, bulk of the $\beta$ decay rates have been studied using theoretical models, primarily because of the lack of experimental data. 

The proton-neutron quasiparticle random phase approximation (pn-QRPA) was employed to compute the Gamow-Teller (GT) strength distributions and $\beta$-decay half-lives~\cite{Mut89}. The pn-QRPA model is widely utilized for a reliable computation of the $\beta$ decay properties of unstable nuclei, both in terrestrial and stellar conditions.
Approximately 6000 nuclei exist between neutron-drip and  stability lines.
The characteristics of $\beta$ decay were investigated  using the pn-QRPA model with a schematic and separable interaction under terrestrial~\cite{Sta90, Hir93, Hom96} and stellar conditions~\cite{Nab99a, Nab04}. More recently, the pn-QRPA model was used to investigate the $\beta$ decay properties of neutron-rich even–even iron (Fe) isotopes with mass numbers $50 \leq A \leq 70$~\cite{Nab24a}.


The properties of $\beta$ decay provide valuable insights for a better understanding of the overall nuclear structure~\cite{Fer34}. Considerable interest has been drawn to the study of nuclear structure due to the evolution of nuclear shapes (e.g.,~\cite{Cej10}). The isotopes of Germanium (Ge) are situated within one of the most intricate sections of the nuclear chart, prompting extensive study of their structure and decay patterns through experimental~(e.g.,\cite{Gur13}) and theoretical (e.g., ~\cite{ Nik14}) approaches. Various theoretical models, such as the shell model~(e.g.,~\cite{Kan15}), the energy density functional framework (e.g.,~\cite{Wang15}),  self-consistent mean-field model (e.g.,~\cite{Ben03}),  interacting boson model (e.g.,~\cite{Die85}) and interacting boson-fermion model (e.g.,~\cite{Iac80}) have been utilized to predict the nuclear structure properties.

Two frequently used nuclear models for determination of nuclear structure properties are the Relativistic Mean Field (RMF) theory and the pn-QRPA model. Both these models have a decent track-record and are known to calculate reliable nuclear data. The RMF theory effectively describes nuclear properties using relativistic dynamics, where nucleons interact through large scalar and vector meson fields. The interplay of these fields explains both the weak average potential and the large spin-orbit splitting. RMF models rely on minimal parameters, fitted to nuclear matter properties and experimental data from nuclei near the $\beta$-stability line~\cite{Ser05}.
The pn-QRPA model is extensively used with both realistic (e.g., \cite{Tan17}) and schematic (e.g., \cite{Hir93}) interactions. A key advantage of using a schematic and separable interaction is that it reduces the problem of Hamiltonian diagonalization to solving a fourth-order algebraic equation. This significantly enhances the computational efficiency,  particularly in calculating Gamow-Teller (GT) strength functions off parent excited states. The latter are required for stellar rate calculations. The pn-QRPA approach employs a separable interaction in a multi-$\hbar \omega$ space, allowing for state-by-state calculation of weak interaction rates. It does so by summing Boltzmann-weighted, microscopically estimated Gamow-Teller (GT) strengths for all parent excited states. This distinctive feature makes it unique compared to other methods for calculating stellar weak rates, including the independent particle model and shell model approaches \cite{Nab19}.

In the present analysis, we have studied the ground-state properties (including the nuclear radii, neutron skin thickness (r$_n$\,-\,r$_p$) and binding energies) of selected Ge isotopes. For this purpose we utilized the RMF framework \cite{Wal94} with  DD-ME2 interaction. The RMF framework describes different ground-state nuclear properties of nuclei using a phenomenological approach. For further description one can see Ref.~\cite{Lal05}.
For the investigation of $\beta$ decay properties (including the Gamow-Teller (GT)
strength distributions, half-lives and stellar weak rates), we employed the pn-QRPA model. This microscopic nuclear  model is widely utilized for a reliable computation of the $\beta$ decay properties of unstable nuclei, both under terrestrial and stellar conditions. The present model-based analysis is compared with previously observed and predicted data.

The format of this document is as follows.  In Section 2, we provide a brief explanation of the RMF and pn-QRPA models to calculate the nuclear structure and $\beta$ decay properties, respectively. Section 3 presents the findings and relevant discussions. Section 4 furnishes summary and conclusions of the current investigation.

\section{Formalism}
The nuclear structure properties of the selected Ge isotopes were investigated using the RMF model. The pn-QRPA model was later used to study the $\beta$ decay properties of the nuclei. Below we present a short description of the two nuclear models. 
\subsection{The RMF framework}
The RMF framework is a theoretical tool used for describing the structure and behavior of atomic nuclei. It is an extension of the mean field theory that takes into account special relativity, unlike the non-relativistic mean field models like the Skyrme model. The model assumes that  nucleons move in atomic nucleus in an effective potential generated by the interactions with other nucleons. The key aspect is the inclusion of relativistic effects, accounting for high velocities and energies of nucleons within the nucleus~\cite{rin96}.

Relativistic mean field theory was initially introduced within the context of quantum field theory. In order to describe the properties of nuclear surface accurately, the density dependence was introduced~\cite{Nab24b}, and the model was called CDFT (covariant density functional theory). Interactions of nucleons in the RMF model are based on exchanging of mesons between nucleons in atomic nuclei. In this model, the saturation mechanism are produced via $\sigma$- and $\omega$-meson fields. The scalar vector $\omega$-meson and $\sigma$-meson fields are related to the repulsive and attractive part of nuclear interaction, respectively. The isovector $\rho$-meson provides asymmetry component in this model.     

Three versions of CDFT are mainly used for calculations based on the consideration of meson-nucleon interaction and self-interaction of mesons \cite{men06}. In the present study, meson-exchange has been taken into account for the calculation of binding energy per nucleon ($BE/A$) and neutron skin thickness (r$_{np}$) of the selected Ge isotopes. In this part, the meson-exchange (ME) version of the RMF model is briefly described.  

In this model, a phenomenological Lagrangian density, consisting of three parts, of the following form was considered for solution
\begin{equation}
	\mathcal{L}=\mathcal{L}_N+\mathcal{L}_m+\mathcal{L}_{int},\label{lagden}
\end{equation}  
 where  $\mathcal{L}_N$ is the free nucleon field while $\mathcal{L}_m$ term covers meson and electromagnetic fields. The term $\mathcal{L}_{int}$ includes both the photon-nucleon  and meson-nucleon interactions. Their explicit forms are gives as 
\begin{equation}
	\mathcal{L}_N=\bar{\Psi}(i\gamma_\mu\partial^\mu-m)\Psi,\label{lagnuc}
\end{equation}
where $\psi$ and $m$ are the Dirac spinor and the nucleon mass, respectively.
\begin{eqnarray}
	\mathcal{L}_m &=\frac{1}{2}\partial_\mu\sigma\partial^\mu\sigma-\frac{1}{2}m_\sigma^2\sigma^2-\frac{1}{2}\Omega_{\mu\nu}\Omega^{\mu\nu}+\frac{1}{2}m^2_\omega\omega_\mu\omega^\mu \nonumber \\ 
	&-\frac{1}{4}\overrightarrow{R}_{\mu\nu}.\overrightarrow{R}^{\mu\nu}+\frac{1}{2}m_\rho^2 \overrightarrow{\rho}_\mu.\overrightarrow{\rho}^\mu -\frac{1}{4}F_{\mu\nu}F^{\mu\nu}, \label{lagmes}
\end{eqnarray} 
where arrows indicate isovectors. $m_\sigma$, $m_\omega$ and $m_\rho$, represent the masses of the related mesons while $\Omega_{\mu\nu}$, $\overrightarrow{R}_{\mu\nu}$ and $F_{\mu\nu}$ are field tensors. 
\begin{eqnarray}
	\mathcal{L}_{int}&= -g_\sigma\bar{\Psi}\Psi \sigma - g_\omega\bar{\Psi}\gamma^\mu\Psi\omega_\mu - g_\rho\bar{\Psi}\overrightarrow{\tau}\gamma^\mu\Psi.\overrightarrow{\rho}_\mu \nonumber \\ 
	&-e\bar{\Psi}\gamma^\mu\Psi A_\mu,
	\label{lagrangian}
\end{eqnarray}
where the coupling constants of the related mesons are indicated by $g_\sigma$, $g_\omega$ and $g_\rho$. 

Usage of the Lagrangian density yields the following Hamiltonian density for the static case~\cite{rin96}
\begin{eqnarray}
	\mathcal{H}({\bf r}) &=\Sigma_i^\dagger(\bm{\alpha p}+\beta m)\Psi_i \nonumber\\
	&+ \frac{1}{2}\left[ ({\bf \nabla}\sigma)^2 + m_\sigma^2\sigma^2\right] - \frac{1}{2}\left[ ({\bf \nabla}\omega)^2 + m_\omega^2\omega^2\right]\nonumber \\ 
	&-\frac{1}{2}\left[ ({\bf \nabla}\rho)^2 + m_\rho^2\rho^2\right] - \frac{1}{2}({\nabla \bf A})^2 \nonumber\\
	&+ \left[ g_\sigma\rho_s\sigma+g_\omega j_\mu\omega^\mu + g_\rho \overrightarrow{j}_{\mu}.\overrightarrow{\rho}^{\mu} + ej_{p\mu}A^{\mu}  \right], \label{hamden}
\end{eqnarray} 
where $\rho_s({\bf r})$, $j_\mu({\bf r})$, $\overrightarrow{j}_\mu({\bf r})$ and $j_{p\mu}({\bf r})$ indicate the isoscalar-scalar density, isoscalar-vector, isovector-vector and electromagnetic currents, respectively. Their explicit forms are given as      


\begin{equation}
	\rho_s({\bf r})=\Sigma_{i=1}^A \bar{\Psi}_i({\bf r}) \Psi_i({\bf r}), \label{isoscalar}
\end{equation}

\begin{equation}
	j_\mu({\bf r})=\Sigma_{i=1}^A \bar{\Psi}_i({\bf r}) \gamma_\mu \Psi_i({\bf r}), \label{isovector}
\end{equation}

\begin{equation}
	\overrightarrow{j}_\mu({\bf r})=\Sigma_{i=1}^A \bar{\Psi}_i({\bf r}) \overrightarrow{\tau}\gamma_\mu \Psi_i({\bf r}), \label{isovectorvec}
\end{equation}

\begin{equation}
	j_{p\mu}({\bf r})=\Sigma_{i=1}^Z \bar{\Psi}_i^\dagger({\bf r})\gamma_\mu \Psi_i({\bf r}). \label{electromagnetic}
\end{equation}
By integrating the Hamiltonian density and using the no-sea approach to add in these densities, the total energy was determined
\begin{equation}
	E_{RMF}\left[\Psi,\bar{\Psi},\sigma,\omega^{\mu},\overrightarrow{\rho}^{\mu}, A^{\mu} \right]=\int d^3r \mathcal{H}({\bf r}). \label{toten}
\end{equation}     
The energy density functional variation specified in Eq.~(\ref{toten}) led to the Dirac equation
\begin{equation}
	\widehat{h}_D \Psi_i=\epsilon_i\Psi_i, \label{diraceq}
\end{equation} 
and a set of differential equations given by 
\begin{equation}
	\left[ -\Delta + m_\sigma^2 \right]\sigma=-g_\sigma\rho_s , \label{helm1}
\end{equation} 
\begin{equation}
	\left[ -\Delta + m_\omega^2 \right]\omega^\mu=g_\omega j^\mu , \label{helm2}
\end{equation}
\begin{equation}
	\left[ -\Delta + m_\rho^2 \right]\overrightarrow{\rho}^\mu=g_\rho \overrightarrow{j}^m , \label{helm3}
\end{equation}
\begin{equation}
	-\Delta A^\mu = ej_p^\mu. \label{helm4}
\end{equation} 
The Dirac Hamiltonian ($\widehat{h}_D$) defined in Eq.~(\ref{diraceq}) can be written as
\begin{equation}
	\widehat{h}_D=\bm{\alpha}(\bm{p-\Sigma}) + \Sigma_0 + \beta(m+\Sigma_s). \label{dirham}
\end{equation} 
The explicit forms of the single nucleon self energies represented by $\Sigma$ in Eq.~(\ref{dirham}) are given as
\begin{eqnarray}
	\Sigma_s({\bf r}) &=g_\sigma\sigma ({\bf r}), \\ 
	\Sigma_\mu ({\bf r}) &=g_\omega\omega_\mu ({\bf r}) + g_\rho\overrightarrow{\tau}.\overrightarrow{\rho}_\mu({\bf r}) + eA_\mu({\bf r})\nonumber\\ &+ \Sigma_\mu^R({\bf r}). \label{selfener}
\end{eqnarray}
The following might be used to describe how rearrangement contributes to the vector self-energy
\begin{equation}
	\Sigma_\mu^R=\frac{j_\mu}{\rho_v}\left( \frac{\partial g_\sigma}{\partial \rho_v}\rho_s\sigma + \frac{\partial g_\omega}{\partial \rho_v} j_v\omega^v + \frac{\partial g_\rho}{\partial \rho_v}\overrightarrow{j}_v.\overrightarrow{\rho}^v \right). \label{rear}
\end{equation}    
There is no current for the solution of even-even atomic nuclei. Therefore, the spatial components of the meson field vanish. The Dirac equation Eq.~(\ref{diraceq}) finally transforms to the following equation

\begin{equation}
	\left\lbrace -i{\bm \alpha\nabla}+\beta M^*({\bf r}) + V({\bf r})\right\rbrace \Psi_i({\bf r})=\epsilon_i \Psi_i({\bf r}) , \label{mass1}
\end{equation}
where $M^*({\bf r})$ represents the effective mass and  $V({\bf r})$ stands for the vector potential.
 Eq.~(\ref{rear}) can further be reduced as follows 
\begin{equation}
	\Sigma_0^R=\frac{\partial g_\sigma}{\partial \rho_v}\rho_s\sigma + \frac{\partial g_\omega}{\partial \rho_v} \rho_v\omega + \frac{\partial g_\rho}{\partial \rho_v}\rho_{tv}\rho, \label{rear2}
\end{equation} 
where the isovector density is denoted by $\rho_{tv}$. 
In this study, the density-dependent meson-exchange DD-ME2 functional~\cite{lal05} for RMF model has been used to calculate $BE/A$ and  r$_{np}$ for the selected Ge isotopes.
\subsection{The pn-QRPA Model}
The model was utilized to compute the GT strength distributions and  corresponding $\beta$ decay half-lives. The Hamiltonian was selected of the following form
\begin{equation} \label{eq22}
	\mathscr{H}^{QRPA} = \mathscr{H}^{sp}+\mathscr{V}^{pair }+ \mathscr{V}_{GT}^{ph}+\mathscr{V}_{GT} ^{pp},
\end{equation}
where the first term on the right hand side of Eq.~\ref{eq24}, $\mathscr{H}^{s p}$,  denotes the single-particle Hamiltonian and $\mathscr{V}^{\text{pair}}$ refers to the nucleon-nucleon pairing interaction. The symbols $\mathscr{V}_{G T}^{p p}$ and $\mathscr{V}_{G T}^{p h}$ stand for particle-particle ($pp$) and particle-hole ($ph$) GT interactions, respectively. The single-particle wave functions and energies were determined using the Nilsson model~\cite{Nil55}. The oscillator constant was calculated using the relation   $\hbar\omega=\left(45/\textit{A}^{1/3}-25/\textit{A}^{2/3}\right)$. Similar values were chosen for protons and neutrons. The Nilsson-potential parameters were chosen from Ref.~\cite{Rag84}. 

The pairing force was calculated using the BCS approximation. The calculations were performed separately for protons and neutrons. We used a constant pairing force with strengths ($G_p$ for protons and $G_n$ for neutrons.)
\begin{eqnarray}\label{eq23}
	\mathcal{V}^{pair}=-G\sum_{jkj^{'}k^{'}}(-1)^{l+j-k}s^{\dagger}_{jk}s^{\dagger}_{j-k}
	\nonumber\\~~~~~~~~~~~~~~~~~~~~~~\times(-1)^{l^{'}+j^{'}-k^{'}} s_{j^{'}-k^{'}}s_{j^{'}k^{'}},
\end{eqnarray}
where  $l$ is orbital angular momentum, $k$ is the projection of $l$ on the symmetry axis. The summation over $k$ and $k{'}$ was restricted to positive values. $s^{\dagger}$ is the particle creation operator in spherical basis.
The proton-neutron residual interactions took place through the \textit{pp} and \textit{ph} GT forces in our nuclear model, which were characterized by  interaction constants  $\kappa$ and $\chi$, respectively in the pn-QRPA framework. In order to calculate the $ph$ GT force, we used the equation
\begin{equation}\label{eq24}
	~ ~ ~ ~ ~ ~ ~ ~ ~ ~ ~ ~	\mathcal{V}_{GT}^{ph}= +2\chi\sum^{1}_{\mu= -1}(-1)^{\mu}\mathcal{U}_{\mu}\mathcal{U}^{\dagger}_{-\mu},
\end{equation}
with
\begin{equation}\label{eq25}
	\mathcal{U}_{\mu}= \sum_{j_{p}k_{p}j_{n}k_{n}}<j_{p}k_{p}\mid
	\tau_\pm ~\sigma_{\mu}\mid
	j_{n}k_{n}>s^{\dagger}_{j_{p}k_{p}}s_{j_{n}k_{n}}.
\end{equation}
The $pp$ GT force also took a separable form
\begin{equation}\label{eq26}
	~ ~ ~ ~ ~ ~ ~ ~ ~ ~ ~ ~	\mathcal{V}_{GT}^{pp}= -2\kappa\sum^{1}_{\mu=-1}(-1)^{\mu}\mathcal{O}^{\dagger}_{\mu}\mathcal{O}_{-\mu},
\end{equation}
with
\begin{eqnarray}\label{eq27}
	\mathcal{O}^{\dagger}_{\mu}= \sum_{j_{p}k_{p}j_{n}k_{n}}<j_{n}k_{n}\mid
	(\tau_\pm \sigma_{\mu})^{\dagger}\mid
	j_{p}k_{p}>\nonumber\\~~~~~~~~~~~~~~~~~~~~~~\times (-1)^{l_{n}+j_{n}-k_{n}}s^{\dagger}_{j_{p}k_{p}}s^{\dagger}_{j_{n}-k_{n}},
\end{eqnarray}
where $\tau_{\pm}$ and $\sigma$ represent the isospin raising/lowering and spin operators, respectively. Other symbols have their usual meanings. 

The $Q_{\beta}$ values were calculated using the mass excess values taken from Ref.~\cite{Kon21}. We used nuclear deformation ($\beta_{2}$) values adopted from the finite range droplet model (FRDM) model~\cite{Mol16}. The $\kappa$ and $\chi$ interaction strengths were determined using the relation $0.58/A^{0.7}$ and $5.2/A^{0.7}$, respectively, taken from Ref.~\cite{Hom96}. The model-independent Ikeda sum rule~\cite{Ike63} was obeyed by our model calculation.

The QRPA phonons were used to described the GT transitions as
\begin{equation}\label{eq28}
	A^{\dagger}_{\omega}(\mu)=\sum_{xy}[M^{xy}_{\omega}(\mu)q^{\dagger}_{x}q^{\dagger}_{\overline{y}}-N^{xy}_{\omega}(\mu)q_{y}q_{\overline{x}}].
\end{equation}
The indices $x$ and $y$, in Eq.~\ref{eq28}, represent the terms  $m_{p}\alpha_{p}$ and $m_{n}\alpha_{n}$, respectively. The sum in Eq.~\ref{eq28} was performed over all pairs of proton-neutron satisfying the conditions for allowed transitions: $\mu=m_{p}-m_{n}$ = 0, $\pm$1 and $\pi_{p}.\pi_{n}$=1 ($\pi$ is the parity of the states). The symbol $q$ ($q^{\dagger}$) denotes annihilation (creation) operator of the quasiparticles. These operators later appeared in the RPA equation and $M^{xy}_{\omega}$ ($N^{xy}_{\omega}$) are the forward and backward amplitudes and further see in Ref.\cite{Mut89}.

Next, we computed the quasiparticle transitions. The detailed recipe for the phonon-related multi quasiparticle (odd A nuclei) of current model (pn-QRPA) is given in Ref. \cite{Mut92}. Further, pn-QRPA model extended for the single and multi quasiparticle by extend the RPA equation. The RPA extend for excitation from ground ($J^\pi = 0^+$) state of an even-even nucleus. The parent nucleus has odd nucleon. The one-quasiparticle phonon-correlated stats are define as
\begin{eqnarray}
	|\xi_{corr}\rangle~=~s^\dagger_{\xi}|-\rangle +& \sum_{\zeta, \omega}a^\dagger_{\nu}A^\dagger_{\omega}(\mu)|-\rangle \nonumber~\langle-|[s^\dagger_{\zeta}A^\dagger_{\omega}(\mu)]^{\dagger}H_{31}s^\dagger_{\xi}|-\rangle \nonumber \\
	&\times E_{\xi}(\zeta,\omega)~~~,
	\label{eq29}
\end{eqnarray}
\begin{eqnarray}
	|\zeta_{corr}\rangle~=~s^\dagger_{\zeta}|-\rangle +& \sum_{\xi, \omega}a^\dagger_{\xi}A^\dagger_{\omega}(-\mu)|-\rangle \nonumber~\langle-|[s^\dagger_{\xi}A^\dagger_{\omega}(-\mu)]^{\dagger}H_{31}s^\dagger_{\zeta}|-\rangle \nonumber \\
	&\times E_{\zeta}(\xi,\omega)~~~,
	\label{eq30}
\end{eqnarray}
with
\begin{equation}
	E_{s}(b,\omega)= (\epsilon_{m}-\epsilon_{n}-\omega)^{-1}~~~~~~~m, n = \xi, \zeta ~~~~,
	\label{eq31}
\end{equation}
In Eqs. \ref{eq29} and \ref{eq30} first term presented the protons (neutron) q.p state. The next second terms shows the RPA phonon-correlation admixed with q.p phonon-coupled Hamiltonian.

This Hamiltonian is denoted with $H_{31}$ and defined as 
\begin{eqnarray}
	H_{31}=\sum V_{\xi \zeta,\bar{\xi^{\prime}}\zeta^{\prime}}(x_{\xi}y_{\zeta}x_{\xi^{\prime}}y_{\zeta^{\prime}}-x_{\xi}y_{\zeta}x_{\xi^{\prime}}y_{\zeta^{\prime}}) (s^\dagger_{\xi}a^\dagger_{\zeta}s^\dagger_{\xi^{\prime}}s_{\zeta^{\prime}} + h.c) + \nonumber \\
	\sum V_{\xi \zeta,\xi^{\prime}\bar{\zeta^{\prime}}}(y_{\xi}y_{\zeta}y_{\xi^{\prime}}x_{\zeta^{\prime}}-x_{\xi}x_{\zeta}x_{\xi^{\prime}}y_{\zeta^{\prime}}) (s^\dagger_{\xi}s^\dagger_{\zeta}s^\dagger_{\zeta^{\prime}}s_{\xi^{\prime}} + h.c) ~~~,
	\label{eq32}
\end{eqnarray}     
The term $h.c$ in Eq. \ref{eq32} shows the Hermitian conjugate. The q.p phonon-coupled Hamiltonian based on the Bogoliubov transformation of separable $pp$ and $ph$ GT interaction forces. The summation in Eq. \ref{eq32} satisfying the condition $m_{\pi}-m_{\nu}=\mu$ with $\pi_{\xi}\pi_{\zeta}=1$. The summation applies to all neuron (proton) q.p and phonon states. For the complete derivation of q,p transitions amplitude of correlated states see Ref. \cite{Mut89}.
The reduced GT transition probabilities were computed using the relation
\begin{equation}\label{eq33}
	~ ~ ~ ~ ~ ~ ~ ~ 	{B}_{GT} (\omega) = |\langle \omega, \mu ||\tau_{\pm} \sigma_{\mu}||QRPA \rangle|^2,
\end{equation}
where $\omega$ is the energy eigenvalue (obtained from solution of the RPA matrix equation) and $\mu$ assumed values 0, $\pm$ 1 (selection rule for allowed transitions). The partial half-lives $t_{1/2}$ were calculated using the relation
\begin{eqnarray}\label{eq34}
		t_{1/2} = \frac{C}{(g_A/g_V)^2f_A(A, Z, E){B}_{GT}(\omega)+f_V(A, Z, E){{B}_F(\omega)}},
	\end{eqnarray}
where the value of constant $C$ (= ${2\pi^3 \hbar^7 ln2}/{g^2_V m^5_ec^4}$)  was adopted as 6143 $s$~\cite{Har09}. The ratio  $g_A/g_V$ was taken as -1.2694 \cite{Nak10} and $E$ = ($Q$ - $\omega$), where $Q$ represent the Q-value of the reaction and was computed from Ref.~\cite{Kon21}. $f_V(A, Z, E)$ and $f_A(A, Z, E)$ are the phase space integrals for vector transitions and axial vector transitions, respectively. ${B}_{GT}$ (${B}_{F}$) are the reduced GT (Fermi) transition probabilities. Calculation of reduced Fermi transitions are relatively simple and can be seen from Ref.~\cite{Hir91}. 
	The  half-lives  were calculated by summing over the inverses of partial half-lives and taking the inverse
	\begin{equation}\label{eq35}
		~ ~ ~ ~ ~ ~ ~ ~ ~ ~ ~ ~ ~	T_{1/2} = \left[\sum_{0 \le \omega \le Q} (\frac{1}{t_{1/2}})\right]^{-1}.
	\end{equation}
 For  further details on solution of Eq.~(\ref{eq24}) we refer to ~\cite{Sta90, Hir93, Hir91, Mut92}.
	
The stellar rates from parent excited level $m$ to daughter state $n$ were calculated using
	\begin{equation} \label{eq36}
		~ ~ ~ ~ ~ ~ ~ ~ ~ ~ ~ ~ ~	\lambda _{mn}^{\beta^{\pm} /EC/PC} =\ln 2\frac{f_{mn}^{\beta^{\pm} /EC/PC}(\rho,T
			,E_{f})}{(ft)_{mn}},
	\end{equation}
where the superscripts identify the various kinds of decay mode. $\beta^{\pm}$ denotes positron and electron emission while $EC~ (PC)$ stands for electron (positron) capture rates. 
	The  $(ft)_{mn}$ values are related to the reduced transition probabilities of GT  and Fermi transitions
	\begin{equation} \label{eq37}
		~ ~ ~ ~ ~ ~ ~ ~ ~ ~ ~ ~ ~ 	(ft)_{mn} =C/{B}^{mn},
	\end{equation}
	where
	\begin{equation} \label{eq38}
		~ ~ ~ ~ ~ ~ ~ ~ ~ ~ ~ ~ ~	{B}^{mn}=(g_{A}/g_{V})^{2} {B}_{GT}^{mn} + {B}_{F}^{mn},
	\end{equation}
	\begin{equation} \label{eq39}
		~ ~ ~ ~ ~ ~ ~ ~ ~ ~ ~ ~ ~	{B}_{GT}^{mn} = \frac{1}{2J_{m} +1} \langle{n}\parallel\sum\limits_{k}
		\tau_{\pm}^{k}\overrightarrow{\sigma}^{k}\parallel {m}\rangle|^{2},
	\end{equation}
	\begin{equation} \label{eq40}
		~ ~ ~ ~ ~ ~ ~ ~ ~ ~ ~ ~ ~	{B}_{F}^{mn} = \frac{1}{2J_{m} +1} \langle{n}\parallel\sum\limits_{k}
		\tau_{\pm}^{k}\parallel {m}\rangle|^{2}.
	\end{equation}
		 Construction of low-lying excited levels and calculation of  nuclear matrix elements in our model can be seen from Ref.~\cite{Mut92} and are not reproduced here for space considerations.  The phase space  integrals ($f_{mn}$) are functions of core temperature ($T$), core density ($\rho$) and Fermi energy ($E_f$). For decay reactions, they were computed using  
\begin{equation} \label{eq41}
		f_{mn}^{\beta^{\mp}} = \int _{1 }^{E_\beta}E_k\sqrt{E_k^{2} -1}(E_{\beta} -E_k)^{2} F(\pm
		Z, E_k) (1-\mathcal{R}_{\mp})dE_k,
	\end{equation}
where lower sign was employed for positron and upper sign for electron emissions.  Phase space integrals ($f_{nm}$) for electron  (indicated by upper signs) and positron capture (indicated by lower signs) were computed using
	\begin{equation} \label{eq42}
		f_{mn}^{EC/PC} = \int _{E_l }^{\infty}E_k\sqrt{E_k^{2} -1}(E_{\beta} +E_k)^{2} F(\pm
		Z, E_k)\mathcal{R}_{\mp}dE_k,
	\end{equation}
where $E_k$ is the electron's kinetic energy including rest mass and we have used natural units ($\hbar=m_{e}=c=1$). $E_l$ denotes the total capture threshold energy for electron capture. The Fermi functions, $F (+Z, E_k)$, were computed as in Ref.~\cite{Gov71}. 
	The total $\beta$ decay energy was calculated using
	\begin{equation} \label{eq43}
		~ ~ ~ ~ ~ ~ ~ ~ ~ ~ ~ ~ ~	E_{\beta} = m_{p} -m_{d} + E_{m} -E_{n},
	\end{equation}
	where $E_{n}$ and $m_{d}$ are the excitation energies and mass of daughter nucleus while, $E_{m}$ and $m_{p}$ represent the corresponding quantities of parent nucleus, respectively. 
	The  distribution functions were computed using  
	\begin{equation} \label{eq44}
		~ ~ ~ ~ ~ ~ ~ ~ ~ ~ ~ ~ ~	\mathcal{R}_{-} =\left[\exp \left(\frac{E_k-E_{f} }{k_{\beta}T} \right)+1\right]^{-1},
	\end{equation}
	\begin{equation} \label {eq45}
		~ ~ ~ ~ ~ ~ ~ ~ ~ ~ ~ ~ ~	\mathcal{R}_{+} =\left[\exp \left(\frac{E_k+2-E_{f} }{k_{\beta}T} \right)+1\right]^{-1}.
	\end{equation}
	where $k_{\beta}$ is the Boltzmann constant.
	
The electron number density, corresponding to protons and nuclei, was calculated using the relation 
	\begin{equation} \label{eq46}
		\rho Y_{e}N_A=\frac{1}{\pi^2}\left(\frac{m_ec}{\hbar} \right)^3\int_0^\infty(\mathcal{R}_--\mathcal{R}_+)p^2dp,
	\end{equation}
where	$N_{A}$ is Avogadro number, $Y_{e}$ is the ratio of electron number to the baryon number and $p$ is the electron or positron momentum.
	
	The total stellar rates were determined using
	\begin{equation} \label{eq47}
		~ ~ ~ ~ ~ ~ ~ ~ ~ ~ ~ ~ ~	\lambda^{\beta^{\pm} /EC/PC} =\sum _{mn}P_{m} \lambda _{mn}^{\beta^{\pm} /EC/PC},
	\end{equation}
	where $P_m$ stands for the occupation probability of the parent excited state and was determined using the Boltzmann distribution. The summation over initial and final states was carried out until desired convergence level was achieved in our rate calculation.
	
\section{Results and Discussion}
\begin{figure}[H]
\centering
{\includegraphics[scale=.5]{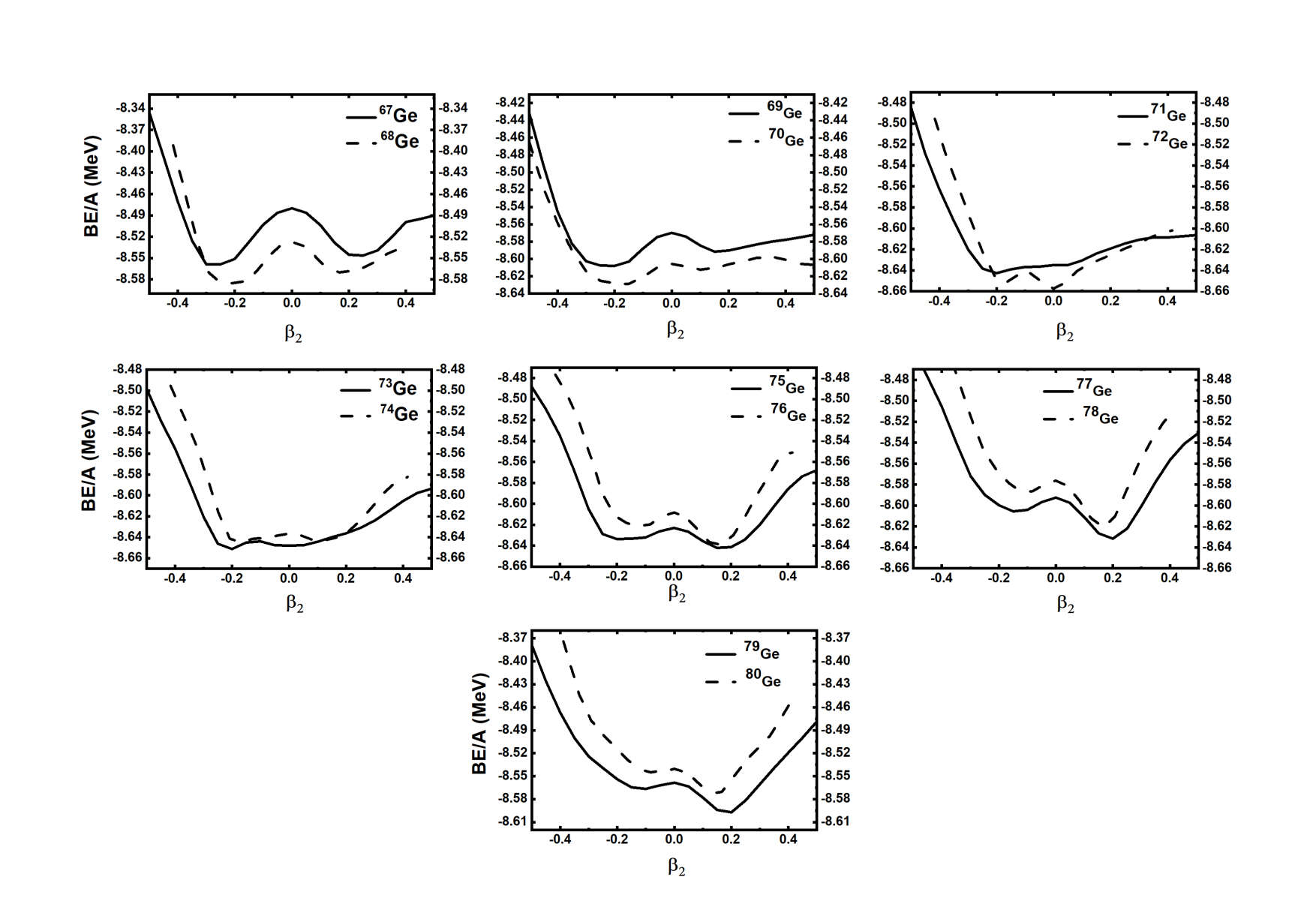}}
\caption{Computed  $BE/A$ of $^{67-80}$Ge nuclei utilizing the RMF framewor with DD-ME2 interaction functional vs $\beta_{2}$. The solid curve represents $BE/A$ for the odd-A Ge isotopes while the dash curve represents $BE/A$ for the even-A Ge isotopes.}
\label{Fig. 01}
\end{figure}

\begin{figure}[H]
	\centering
	\includegraphics[width=0.7\linewidth]{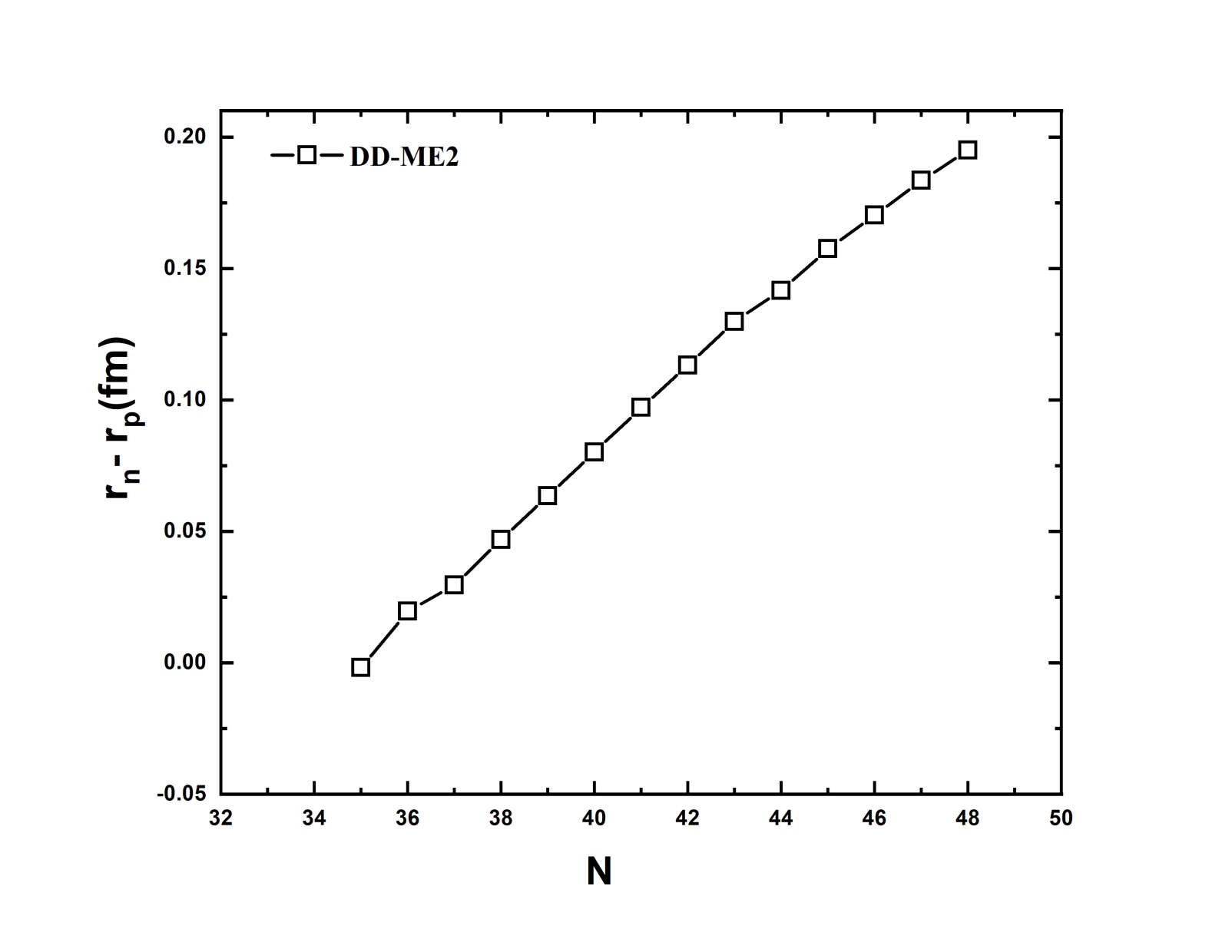}
	\caption{Calculated neutron skin thickness for Ge isotopes within the framework of RMF with DD-ME2  interaction functional.}
	\label{Fig. 02}
\end{figure}

In this study, we examined the ground-state properties (neutron skin thicknesses and binding energies) of the  neutron-rich Ge isotopes within the framework of RMF approach. The $\beta$ decay properties, including log $\textit{ft}$ values, GT strength distributions, half-lives and stellar weak rates, were calculated in a fully microscopic manner using the pn-QRPA model. The RMF model is one of the most effective nuclear models utilized to examine the nuclear ground-state properties. In Fig.~\ref{Fig. 01}, we display the $BE/A$ vs $\beta_{2}$ for $^{67-80}$Ge. For the Ge isotopes, the evolution of the $BE/A$ with $N$ is observed: it evolves from an oblate shape to soft spherical and  then to a prolate shape as $N$ grows.
 We have calculated axially symmetric RMF model calculation, using the density-dependent meson-exchange DD-ME2~\cite{Lal99} functional, for obtaining the ground-state properties of the $^{67-80}$Ge nuclei. In the present investigation, we employed the  $\beta_{2}$ constrained axially symmetric RMF model as provided in \cite{Nik08}, to compute the binding energy and neutron skin thickness for the selected isotopes. The ground-state shape of $^{67-70}$Ge and $^{72, 73}$Ge  are oblate. $^{71}$Ge and $^{74}$Ge have smooth spherical shapes while $^{75-80}$Ge have almost prolate shapes, as per prediction of the DD-ME2 interaction. Comparison of our calculated $BE/A$ verses the experimental data \cite{Kon21} are depicted in Table.~\ref{Tab 1}. The $BE/A$ of the selected nuclei displays a minimum near $\beta_{2}$=0.2. The discrepancies from the available experimental data suggest that new degrees of freedom, e.g., triaxial deformations, are required in the model which we plan to include as a future assignment.
 A correlation has been observed between the thickness of the neutron skin and the slope parameter of the nuclear matter symmetry energy. $r_{np}$ can be defined using the root mean square (rms) radii of neutrons $r_{n}$ and protons $r_{p}$. Fig.~\ref{Fig. 02} illustrates the computed neutron skin thickness of $^{67-80}$Ge  employing the RMF framework with DD-ME2 interaction. As widely understood from the knowledge of fundamental nuclear physics, neutron skin thickness increases for N $\gtrsim$ Z. This results due to the fact that neutrons are adding in the isotopic chain of nuclei. {However, for N $\sim$ Z, r$_{np}$ $\sim$ 0 and this particular scenario is obvious for $^{67}$Ge}. 
\begin{threeparttable}[H]
	\centering
	\caption{Comparison of calculated binding energy per nucleon with measured data~\cite{Kon21}.} 
	\label{Tab 1}
	\begin{tabular}{|c c c| c c c|}
		\hline 
		
		~~~~~~~~ & \textbf{Odd-A} & ~~~~~~~~ &~~~~~~~~ & \textbf{Even-A} & ~~~~~~~~ \\ \hline
		 
		\textbf{Nuclei} & \textbf{$$BE/A$$ \cite{Kon21}} & \textbf{$$BE/A$$ (This work)}&\textbf{Nuclei} & \textbf{$$BE/A$$ \cite{Kon21}} & \textbf{$$BE/A$$ (This work)} \\ \hline   
		$^{67}$Ge& 8.63& 8.57&$^{68}$Ge & 8.68& 8.59\\ 
		$^{69}$Ge& 8.68& 8.61&$^{70}$Ge & 8.72& 8.64 \\
		$^{71}$Ge& 8.70& 8.65&$^{72}$Ge & 8.73& 8.66 \\
		$^{73}$Ge& 8.71& 8.66&$^{74}$Ge & 8.72& 8.65 \\ 
		$^{75}$Ge& 8.69& 8.64&$^{76}$Ge & 8.70& 8.64 \\ 
		$^{77}$Ge& 8.67 & 8.62&$^{78}$Ge & 8.67& 8.62 \\
		$^{79}$Ge& 8.63& 8.60&$^{80}$Ge & 8.62& 8.58 \\ 
		   \hline
	\end{tabular}
\end{threeparttable}
\\

\begin{figure}[H]
	\centering
	{\includegraphics[scale=.5]{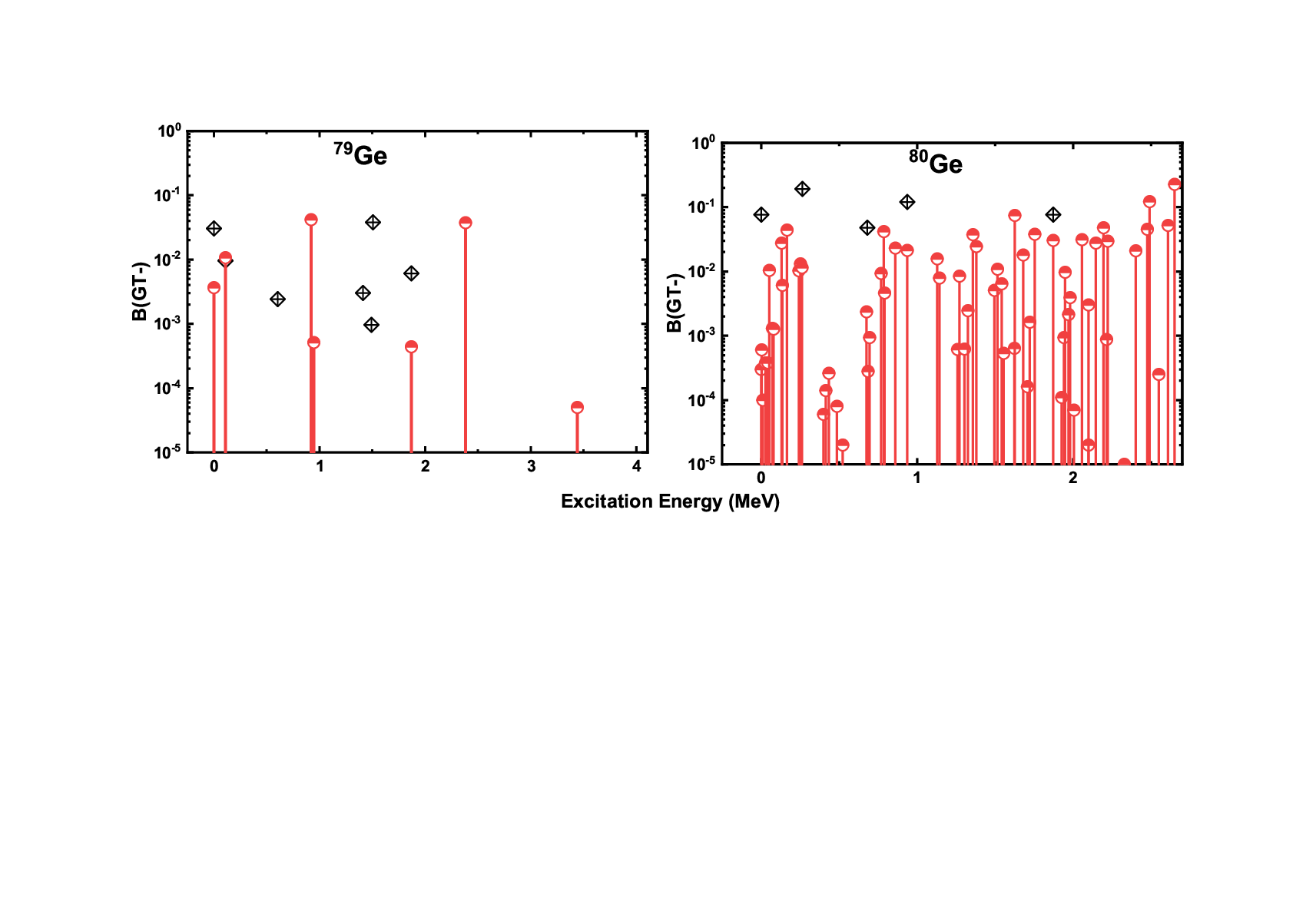}}
	\caption{Comparison of the pn-QRPA calculated GT strength distributions with experimental data~\cite{Hof81}. The excitation energies are given in daughter nuclei and stated up to $Q_{\beta}$ window.}
	\label{Fig. 3}
\end{figure}

The $\beta$ decay properties of Ge isotopes were determined using the pn-QRPA model. We start the proceedings by comparing our calculated GT distributions with measured data to validate our nuclear model. 
Fig.~\ref{Fig. 3} illustrates the comparison between our computed GT strength functions for $^{79}$Ge and $^{80}$Ge and the measured values. The measured data was taken from Ref.~\cite{Hof81}. The experimental data was reported up to incident energy  E$_x$= 1.87~ MeV. The calculated GT transitions are shown as a function of daughter excited levels within the $Q$-window. For the case of $^{79}$Ge and $^{80}$Ge our model reproduces well the low-lying transitions. It is however noted that, for $^{79}$Ge, our model misses few low-lying transitions. Our model incorporated residual correlations among nucleons through one-particle one-hole (1p-1h) excitations within a spacious multi-7$\hbar \omega$ model space.  Inclusion of 2p-2h or higher excitations may shift the strength to low-lying states in daughter. Overall a decent agreement between measured and calculated GT strength distributions can be noted from Fig.~\ref{Fig. 3}.

Table~\ref{Tab 2} depicts the decay modes and $Q_{\beta}$ values of the selected Ge isotopes determined from the NUBASE2020 evaluation~\cite{Kon21}. The  $Q_{\beta}$ values were computed using the mass excess given in Column~2 of Ref.~\cite{Kon21}. Negative computed values of $Q_{\beta}$  imply that the decay is not spontaneous. For $^{76}$Ge, the NUBASE2020 evaluation shows a half-life of 1.88 $\times$ 10$^{21}$ years with double $\beta$ decay mode. The same data is shown in Table~\ref{Tab 2}.
 We calculate  log $ft$ values, GT distributions and half-lives according to the spontaneous decay modes shown in Table~\ref{Tab 2}. 

Table~\ref{Tab 3} shows the comparison of calculated log $\textit{ft}$ values with measured data and previous calculation using the  proton-neutron interacting boson-fermion model (IBFM-2)~\cite{Nom22}  for the state-by-state transitions of odd-A Ge nuclei. The pairing interaction and a universal energy density functional (EDF) were used in constrained self-consistent mean-field calculations, to determine the interacting-boson Hamiltonian for the even-even core nuclei, the essential components of the particle-boson interactions for the odd-nucleon systems, and of the charge-changing transition operators. The experimental data was taken from~\cite{NNDC} and shown in column 4 of Table~\ref{Tab 3}. The pn-QRPA calculated values are generally bigger than the IBFM-2 data except for the higher lying 1/2$^-$ and 3/2$^-$ states in $^{75}$As. The IBFM-2 calculated log $ft$ values are too sensitive to the initial and final wave functions~\cite{Nom22}. The pn-QRPA log $ft$ values are observed to be in better consistency with the measured values. The largest deviation occurs, for example, in the case of the $^{75}$Ge($1/2^+_1$) $\rightarrow$ $^{75}$As($3/2^+_2$) decay. In particular, the computed log $\textit{ft}$  is a factor of 1.64 larger than the IBFM-2. Here a GT transition is involved. The reduced GT matrix element is found to be as small as 0.00002, due to the cancellation of the many components that make the matrix element.Similarly, Table~\ref{Tab 4} show the comparison of our calculation of  log $\textit{ft}$ values with measured and the proton-neutron interacting boson fermion-fermion model (IBFFM-2) model~\cite{Nom22} for even-even isotope of Ge. The calculated pn-QRPA log $\textit{ft}$ values are larger than measured and IBFFM-2 calculations. In the decay process $^{78}$Ge($0^+_1$) $\rightarrow$ $^{78}$As($1^+_1$), the pn-QRPA log $\textit{ft}$ value is higher than the observed value. This discrepancy occurs because several terms in the GT matrix element are fragmented and cancel each other out, resulting in a relatively large log $\textit{ft}$ value. For the decay $^{78}$Ge($0^+_1$) $\rightarrow$ $^{78}$As($1^+_2$), the comparison is perfect.

\begin{figure}[H]
	\centering
	{\includegraphics[scale=.35]{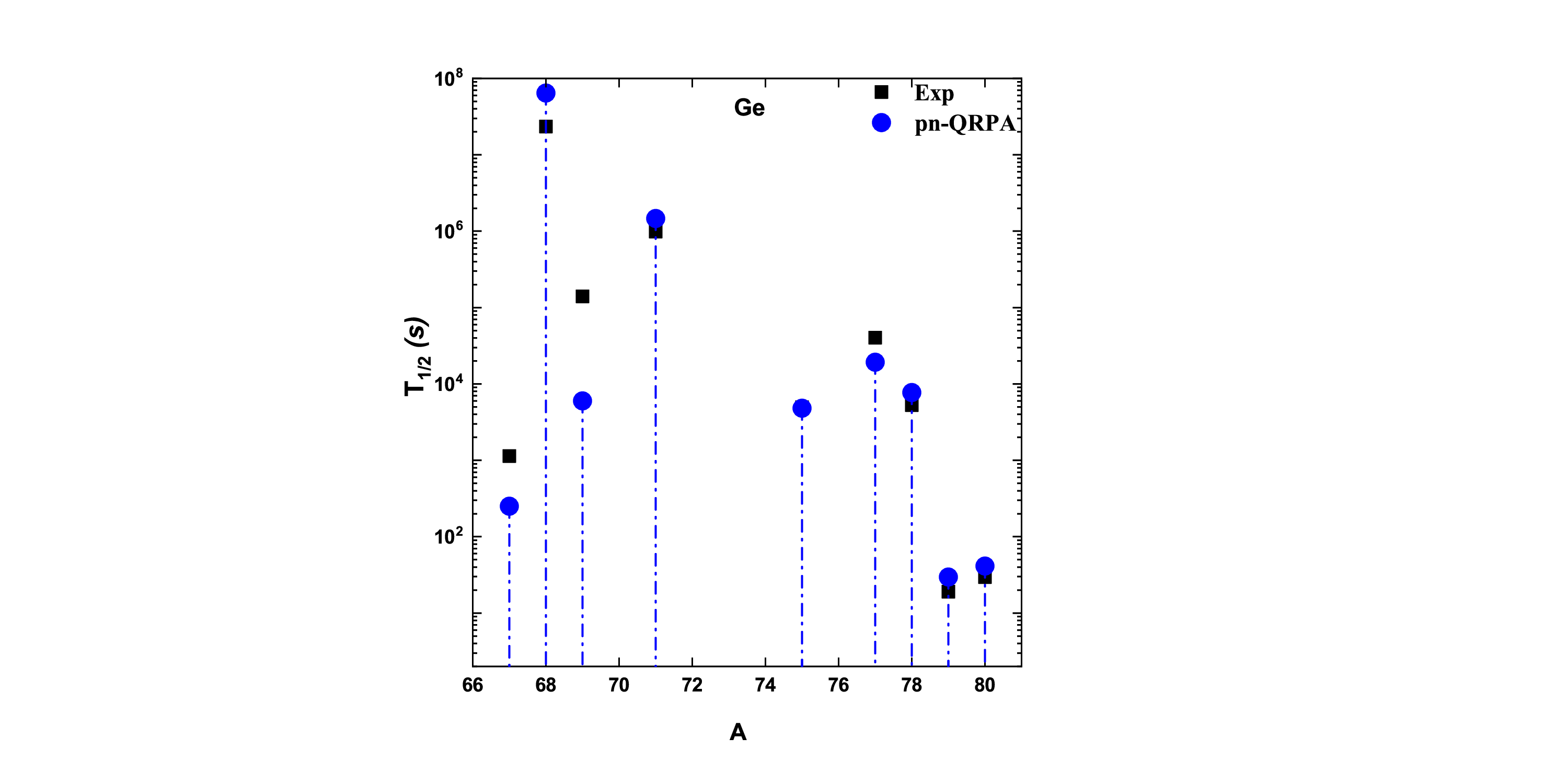}}
	\caption{The pn-QRPA calculated $\beta$ decay half-lives of Ge isotopes compared with the experimental data~\cite{Kon21}.}
	\label{Fig. 4}
\end{figure}

\begin{figure}[H]
	\centering
	{\includegraphics[scale=.4]{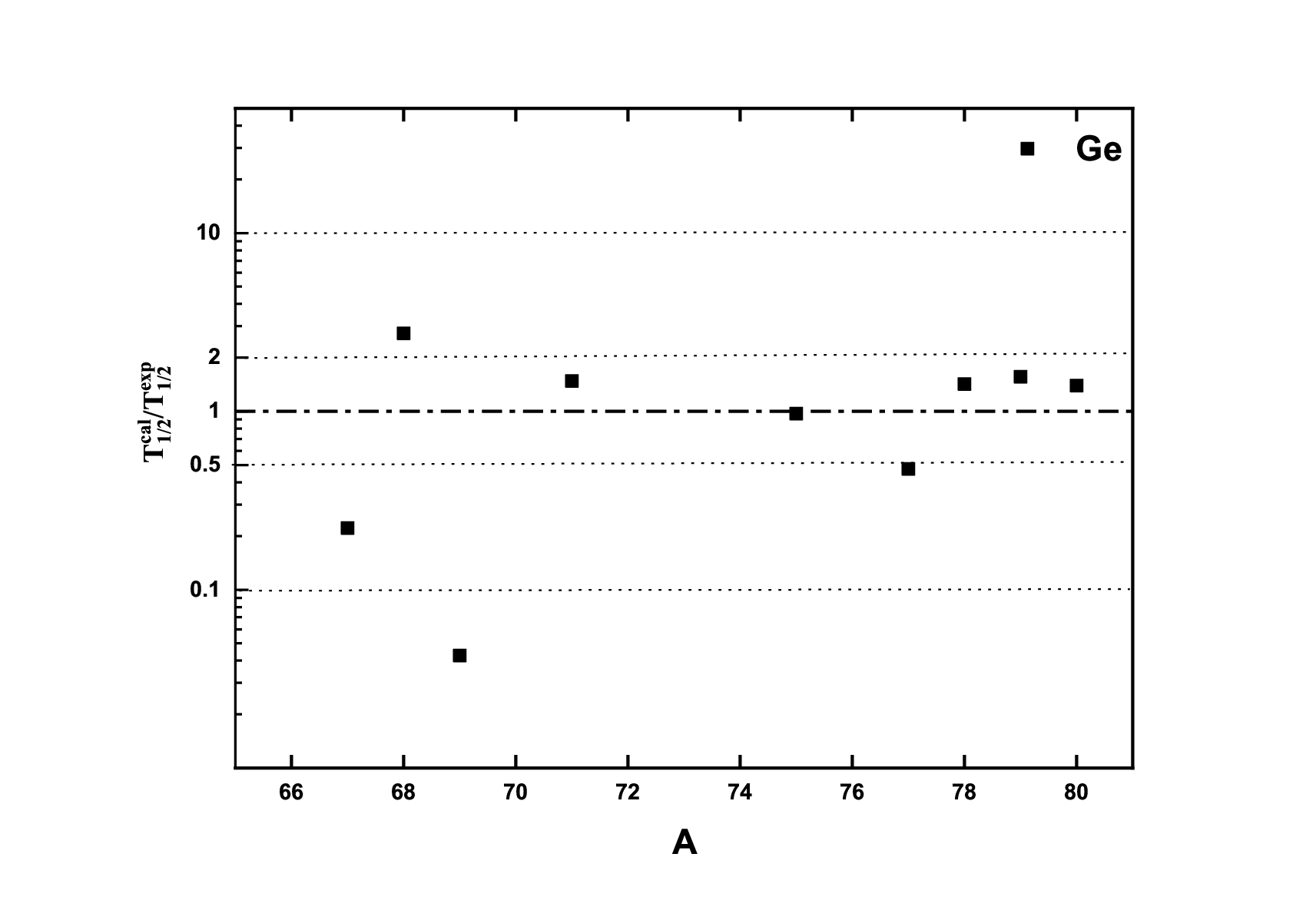}}
	\caption{Ratio of the pn-QRPA calculated half-lives of Ge isotopes to measured data~\cite{Kon21}.}
	\label{Fig. 5}
\end{figure}

Fig.~\ref{Fig. 4} illustrates the comparison between calculated half-lives of Ge across various mass numbers and experimental data.  $^{70}$Ge, $^{72-74}$Ge are stable nuclei whereas $^{76}$Ge is $\beta \beta$ unstable. Fig.~\ref{Fig. 4} shows a good comparison between our computed half-lives and the corresponding experimental data. The measured half-lives for $^{67}$Ge and $^{69}$Ge are 1.13$\times10^{3}$ $s$ and 1.41$\times10^{5}$ $s$ \cite{Kon21}, respectively. The corresponding calculated half-lives are 2.50$\times10^{2}$ $s$ and 2.13$\times10^{4}$ $s$. The calculated half-lives are a factor 4.52 and 6.62, respectively, smaller than the measured data. The computed high branching ratios with their corresponding GT strength distributios at low-lying excitation energies are shown in Table~\ref{Tab 5}. The higher calculated $\beta$-decay rates led to shorter computed half-lives for $^{67,69}$Ge. Fig.~\ref{Fig. 5} displays the ratios of pn-QRPA-calculated half-lives to the measured half-lives of Ge isotopes. The calculated values are  generally reproduced within a factor of 2 of the measured half-lives. The pn-QRPA model is known to have a good prediction power for neutron-rich nuclei~\cite{Hir93,Sta90}. 

We have used our nuclear model to calculate $\beta^{\pm}$, $EC$ and $PC$ rates of selected Ge isotopes in stellar matter. Earlier these stellar  rates were calculated using the independent shell model and the then available experimental data for free nucleons and 226 nuclei in the mass region $A$ = 21--60~\cite{Ful82}. Later Pruet and Fuller~\cite{Pre03} (referred to as PF03 here onward), updated these calculations for  the mass region $A$ = 65--80. In the revised calculation, PF03 remedied the misassignment of placement of GT centroids by Ref.~\cite{Ful82} and introduced a better treatment of high-temperature partition functions. However the theoretical framework of the model was same as used by Ref.~\cite{Ful82}.  We present the pn-QRPA calculated stellar rates for selected Ge isotopes in  Tables~(\ref{Tab 6}~--~\ref{Tab 7}). Here we show the sum of $\beta^{+}$ and $EC$ in one direction and sum of $\beta^{-}$ and $PC$ in the other direction, for the selected Ge isotopes. Tables~(\ref{Tab 6}~--~\ref{Tab 7}) show the comparison of our calculated ($\beta^{+}$+$EC$) and ($\beta^{-}$+$PC$) rates, respectively, with those of PF03 results for Ge isotopes  at selected temperature (3, 10, 30) GK and density ($10^3, 10^7$, $10^{11}$)~g/cm$^3$ values. The decay rates increase as the core temperature raises. This is because the occupation
  probability for parent states increases with the increase in stellar temperature , leading to a significant contribution to the total rates. As the values of $\rho Y_{e}$ rise, the ($\beta^{+}$+$EC$)  rates also increase  due to a corresponding elevation in electron chemical potential. Conversely, the ($\beta^{-}$+$PC$) rates get decrease for high $\rho Y_{e}$ values due to a substantial decrease in the available phase space caused by increased electron chemical potential.
 Table~\ref{Tab 6} shows that the pn-QRPA rates are generally bigger than PF03 rates, up to an order of magnitude, at high core temperature ($T_9$) and density ($\rho Y_{e}$) values. 
Table~\ref{Tab 7} shows that the pn-QRPA ($\beta^{-}$+$PC$) rates are smaller than PF03 rates at small and intermediate core temperatures. Once again at high $T_9$ values, our rates are up to an order of magnitude bigger than PF03 rates. The PF03 method applied a quenching factor of 3 for GT- transitions and 4 for GT+ transitions, respectively. No explicit quenching factor was included in our calculation. The partition function sum, in the PF03 computation, had a large number of states without the associated weak interaction strength. These may be a probable reason for the smaller PF03 rates especially at high $\rho Y_{e}$ and $T_9$ values.

\section{Conclusions}
The nuclear ground state parameters for $^{67-80}$Ge nuclei were computed employing the axially deformed RMF model in the initial phase of our investigation. The results of our computations for the ground-state binding energy per nucleon and  neutron-proton radii,  were reported. In this study, we explored potential relationships between the neutron skin thickness of neutron-rich nuclei and nuclear matter properties. The axially deformed RMF model was utilized to examine the characteristics of the nuclear ground state by utilizing density-dependent DD-ME2 interaction. It was noticed that the computed binding energies and the available experimental data agreed fairly well. The computed $BE/A$ curve of the isotopes predict that the ground-state shape of Ge nuclei, as studied in this project, are not spherical. E.g., the ground-state shapes of $^{(67, 68, 74-76, 79)}$Ge are expected to be prolate and oblate, respectively, using the DD-ME2 interaction. 

We later investigated the $\beta^\pm$ and $EC$/$PC$ rates for Ge nuclei  in the region $67 \leq A \leq 80$. The pn-QRPA framework was used to determine the $\beta$ decay properties of chosen Ge isotopes. The calculated half-life and log $\textit{ft}$ values were in decent agreement with the corresponding measured data. The same nuclear framework was utilized to calculate weak  decay rates of Ge nuclei in stellar environment. We compare our calculated ($\beta^{+}$+EC) and ($\beta^{-}$+PC) rates with the PF03 rates. It was concluded that the pn-QRPA rates are bigger, up to an order of magnitude, than the PF03 rates at high stellar temperature and density values. Quenching of the GT strength and improper handling of partition functions  by PF03 could be the potential reasons for their smaller rates. The pn-QRPA stellar rates were fully microscopic in nature and did not assume the Brink-Axel hypothesis in  calculation of excited states GT distributions.   We hope the reported stellar rates would prove useful for $r$-process nucleosynthesis calculations and simulations of late time stellar evolution.  

\section*{Acknowledgments}  
J.-U. Nabi and A. Mehmood would like to acknowledge financial support of the Higher Education Commission Pakistan through project number 20-15394/NRPU/R\&D/HEC/2021.  

\section*{Declaration of Competing Interest}  

The authors declare that they have no known competing financial interests or personal relationships that could have appeared to
influence the work reported in this paper.
\centering

\begin{threeparttable}[H]
	\centering
	\caption{Terrestrial decay modes of selected Ge isotopes. Q-values were taken from Ref.~\cite{Kon21}.} 
	\label{Tab 2}
	\begin{tabular}{c c c}
		\hline \\
		\textbf{Nuclei} & \textbf{Decay Mode} & $\mathbf{Q_{\beta}(MeV)}$ \\ \hline \\  
		$^{67}$Ge & $\beta^{+}$ &  4.2052  \\  
		$^{68}$Ge & EC & 0.1073 \\  
		$^{69}$Ge & $\beta^{+}$ & 2.2271  \\  
		$^{70}$Ge & Stable & -1.6518\\  
		$^{71}$Ge & EC & 0.2324 \\  
		$^{72}$Ge & Stable & -3.9976  \\  
		$^{73}$Ge & Stable & -0.3445  \\  
		$^{74}$Ge & Stable &  -2.5623  \\  
		$^{75}$Ge & $\beta^{-}$ &  1.1772   \\  
		$^{76}$Ge & 2$\beta^{-}$ & -0.9215  \\  
		$^{77}$Ge & $\beta^{-}$ &  2.7034  \\  
		$^{78}$Ge & $\beta^{-}$ &  0.9550  \\  
		$^{79}$Ge & $\beta^{-}$ &  4.1060  \\  
		$^{80}$Ge & $\beta^{-}$ &  2.6797  \\  
		\\   \hline
	\end{tabular}
\end{threeparttable}

\centering

\centering

\begin{threeparttable}[H]
	\centering
	\caption{ Comparison of calculated and measured~\cite{NNDC} log $\textit{ft}$ values for $\beta^{-}$ decays of odd-A Ge nuclei. The third and last column show the theoretical values calculated using the IBFM-2 \cite{Nom22} and pn-QRPA models, respectively.\\}
	\label{Tab 3}
	\addtolength{\tabcolsep}{1pt}
	\begin{tabular}{ccccccc}
		\hline
		\multicolumn{7}{c}{~ ~ ~ ~ ~ ~ ~ ~ ~ ~ ~ ~ ~ ~ ~ ~ ~ ~ log $\textit{ft}$ }\\
		\hline
		Decay & $I_{i} \rightarrow I_{f}$ & IBFM-2 & Exp. & pn-QRPA & \\
		\hline
		$^{75}$Ge $\rightarrow$ $^{75}$As & $1/2^{-}_1 \rightarrow 3/2^{-}_1$ & 3.54 & 5.17 & 5.08   & & \\
		& $1/2^{-}_1 \rightarrow 3/2^{-}_2$ & 5.04 & 5.63 & 8.28   & & \\
		& $1/2^{-}_1 \rightarrow 3/2^{-}_3$ & 5.90 & 6.42 & 5.36   & & \\
		& $1/2^{-}_1 \rightarrow 1/2^{-}_1$ & 5.66 & 6.87 & 5.68   & & \\
		& $1/2^{-}_1 \rightarrow 1/2^{-}_2$ & 4.45 & 6.94 & 5.08   & & \\
		& $1/2^{-}_1 \rightarrow 1/2^{-}_3$ & 7.18 & 6.42 & 5.07   & & \\
		$^{77}$Ge $\rightarrow$ $^{77}$As & $1/2^{-}_1 \rightarrow 3/2^{-}_1$ & 3.38 & 4.96 & 7.98   & & \\
		& $1/2^{-}_1 \rightarrow 3/2^{-}_2$ & 3.79 & 7.20  & 6.48   & & \\
		& $1/2^{-}_1 \rightarrow 3/2^{-}_3$ & 4.71 & 5.30 & 7.58   & & \\
		& $1/2^{-}_1 \rightarrow 3/2^{-}_4$ & 4.82 & 7.20 & 5.82   & & \\
		& $1/2^{-}_1 \rightarrow 3/2^{-}_5$ & 5.15 & 5.70 & 6.77   & & \\
		& $1/2^{-}_1 \rightarrow 1/2^{-}_1$ & 5.32 & 7.70 & 6.77   & & \\
		& $1/2^{-}_1 \rightarrow 1/2^{-}_2$ & 4.91 & 5.70 & 7.88   & & \\
		& $1/2^{-}_1 \rightarrow 1/2^{-}_3$ & 3.98 & 5.80 & 6.63   & & \\
		\hline
	\end{tabular}
\end{threeparttable}

\vspace{2cm}
\begin{threeparttable}[H]
	\centering
	\caption{Comparison of calculated and measured~\cite{NNDC} log $\textit{ft}$ values for $\beta^{-}$ decay of even-even nucleus $^{78}$Ge. The third and last column show the theoretical values calculated using the  IBFFM-2 \cite{Nom22} and pn-QRPA models, respectively.\\}
	\label{Tab 4}
	\addtolength{\tabcolsep}{1pt}
	\begin{tabular}{ccccccc}
		\hline
		\multicolumn{7}{c}{~ ~ ~ ~ ~ ~ ~ ~ ~ ~ ~ ~ ~ ~ ~ ~log $\textit{ft}$ }\\
		\hline
		Decay & $I_{i} \rightarrow I_{f}$ & IBFFM-2 & Exp. & pn-QRPA & \\
		\hline
		$^{78}$Ge $\rightarrow$ $^{78}$As & $0^+_1 \rightarrow 1^+_1$ & 3.92 & 4.26 & 5.98   & & \\
		& $0^+_1 \rightarrow 1^+_2$ & 5.15 & 5.61 & 5.60   & & \\
		\hline
	\end{tabular}
\end{threeparttable}
\vspace{0.2cm}

\begin{threeparttable}[H]
	\centering
	\caption{The low-lying transitions with branching ratios for 	$^{67,69}$Ge } 
	\label{Tab 5}
	\begin{tabular}{c c c c}
		\hline \\
		\textbf{Nuclei} & \textbf{Excitation Energy (MeV)} & \textbf{${B}_{GT}$ arb. units}& \textbf{Branching Ratios}\\ \hline \\  
		$^{67}$Ge & 0.30 & 0.04442 & 60.82 \\  
		
		$^{69}$Ge & 0.00 & 0.02764 & 34.64 \\  
		
		\\   \hline
	\end{tabular}
\end{threeparttable}

\newpage
\begin{threeparttable}[H]
	\tiny
	\centering
	\caption{Sum of pn-QRPA calculated  stellar $\beta^{+}$ and $EC$ rates  for Ge isotopes as a function of core temperature and density values in units of $s^{-1}$.  Temperatures ($T_9$) are given in units of $10^9$K and densities ($\rho Y_e$) are given in units of g/cm$^3$. The rates are compared with the PF03~\cite{Pre03} calculation.} \label{Tab 6}
	\addtolength{\tabcolsep}{1pt}
	\begin{tabular}{|c|c|cc|cc|cc|}
		\hline
 		\textbf{Nuclei} & \textbf{$T_9$} &\multicolumn{2}{c|}{\textbf{$\rho Y_{e}=10^3$}}   &\multicolumn{2}{c|}{ \textbf{$\rho Y_{e}=10^7$}}  & \multicolumn{2}{c|}{\textbf{$\rho Y_{e}=10^{11}$}}\\  \hline
		~ & & $\lambda_{(\beta^{+}+EC)}^{pn-QRPA}$ & $\lambda_{(\beta^{+}+EC)} ^{PF03}$ & $\lambda_{(\beta^{+}+EC)}^{pn-QRPA}$ & $\lambda_{(\beta^{+}+EC)} ^{PF03}$ & $\lambda_{(\beta^{+}+EC)}^{pn-QRPA}$ & $\lambda_{(\beta^{+}+EC)} ^{PF03}$ \\ \hline
		$^{67}$Ge & 3 & 7.69$\times10^{-3}$ & 3.30$\times10^{-3}$ &
		 3.24$\times10^{-2}$ & 2.40$\times10^{-2}$
		 & 8.43$\times10^{4}$ & 4.72$\times10^{4}$
		  \\ 
		~ & 10 & 3.60$\times10^{-1}$ & 2.75$\times10^{-1}$
		& 4.44$\times10^{-1}$ & 3.33$\times10^{-1}$
		& 1.36$\times10^{5}$ & 4.79$\times10^{5}$
		 \\ 
		~ & 30 & 4.36$\times10^{2}$ & 1.67$\times10^{2}$
		 & 4.40$\times10^{2}$ & 1.76$\times10^{2}$
		 & 4.15$\times10^{5}$ & 8.93$\times10^{4}$
		 \\ \hline
		$^{68}$Ge & 3  & 1.68$\times10^{-5}$ & 3.26$\times10^{-5}$
		 & 7.60$\times10^{-4}$ & 1.38$\times10^{-3}$ &
		  5.62$\times10^{4}$ & 3.48$\times10^{4}$
		  \\ 
		~ & 10 & 8.34$\times10^{-2}$ & 1.28$\times10^{-1}$
		 & 1.04$\times10^{-1}$ & 1.57$\times10^{-1}$
		 & 6.00$\times10^{4}$ & 3.65$\times10^{4}$
		 \\ 
		~ & 30 & 8.71$\times10^{1}$ & 7.77$\times10^{1}$ & 8.79$\times10^{1}$ & 7.84$\times10^{1}$ &
		 8.36$\times10^{4}$& 5.15$\times10^{4}$
		 \\ \hline
		$^{69}$Ge & 3 &  1.65$\times10^{-4}$ & 2.30$\times10^{-4}$
		&  3.17$\times10^{-3}$ & 6.82$\times10^{-3}$
		 &  4.94$\times10^{4}$ & 3.01$\times10^{4}$
		 \\ 
		~ & 10 &  8.95$\times10^{-2}$ & 1.45$\times10^{-1}$
		& 1.11$\times10^{-1}$ & 1.80$\times10^{-1}$
		& 8.89$\times10^{4}$ & 3.02$\times10^{4}$
		\\ 
		~ & 30  & 2.54$\times10^{2}$ & 1.15$\times10^{2}$ &
		 2.56$\times10^{2}$ & 1.16$\times10^{2}$
		 &  3.13$\times10^{5}$ & 5.89$\times10^{4}$
		 \\ \hline
		$^{70}$Ge & 3  & 6.01$\times10^{-8}$  & 3.43$\times10^{-7}$ & 3.05$\times10^{-6}$ & 1.74$\times10^{-5}$
		 	 & 3.61$\times10^{4}$ & 1.79$\times10^{4}$ \\ 
		~ & 10 & 2.02$\times10^{-2}$ & 1.73$\times10^{-2}$
		 & 2.53$\times10^{-2}$ & 2.12$\times10^{-2}$
		  & 3.94$\times10^{4}$ & 1.57$\times10^{4}$ \\ 
		~ & 30 & 5.18$\times10^{1}$  & 3.62$\times10^{1}$
		 & 5.21$\times10^{1}$ & 3.65$\times10^{1}$
		  & 6.32$\times10^{4}$ & 2.49$\times10^{4}$
		   \\ \hline 
		$^{71}$Ge & 3  & 6.15$\times10^{-6}$ & 2.32$\times10^{-5}$
		 &	2.36$\times10^{-4}$ & 8.99$\times10^{-4}$
		  &  5.14$\times10^{4}$ & 1.45$\times10^{4}$
		  \\ 
		~ & 10 & 2.78$\times10^{-2}$ & 4.39$\times10^{-2}$
		 & 3.47$\times10^{-2}$ &  5.49$\times10^{-2}$ & 8.63$\times10^{4}$ & 1.48$\times10^{4}$
		  \\ 
		~ & 30 & 2.07$\times10^{2}$ & 6.49$\times10^{1}$
		& 2.09$\times10^{2}$ & 6.53$\times10^{1}$
		& 3.09$\times10^{5}$ & 3.30$\times10^{4}$
		 \\ \hline
		$^{72}$Ge & 3 & 3.81$\times10^{-11}$ & 4.84$\times10^{-11}$
		& 1.88$\times10^{-9}$ & 2.49$\times10^{-9}$
		 & 2.88$\times10^{4}$ & 7.96$\times10^{3}$ \\
		~ & 10 & 3.28$\times10^{-3}$ & 1.67$\times10^{-3}$
		 &  4.11$\times10^{-3}$ & 2.01$\times10^{-3}$
		  &  3.15$\times10^{4}$ & 8.17$\times10^{3}$	\\ 
		~ & 30 & 3.04$\times10^{1}$ & 1.72$\times10^{1}$
		 & 3.07$\times10^{1}$ & 1.73$\times10^{1}$
		  & 4.91$\times10^{4}$ & 1.38$\times10^{4}$
		   \\ \hline
		$^{73}$Ge & 3 & 6.27$\times10^{-8}$ & 5.06$\times10^{-8}$
		 &  3.13$\times10^{-6}$ & 2.62$\times10^{-6}$ &
		  4.09$\times10^{4}$ & 1.10$\times10^{4}$
		   \\ 
		~ & 10 & 6.00$\times10^{-3}$ & 9.40$\times10^{-3}$
		 & 	 7.52$\times10^{-3}$ & 1.18$\times10^{-2}$
		  & 6.49$\times10^{4}$ & 1.14$\times10^{4}$
		   \\ 
		~ & 30 & 1.35$\times10^{2}$ & 4.14$\times10^{1}$
		 &	1.36$\times10^{2}$ & 4.17$\times10^{1}$
		  & 2.35$\times10^{5}$ & 2.39$\times10^{4}$
		   \\ \hline
		$^{74}$Ge & 3 & 3.47$\times10^{-13}$ & 6.55$\times10^{-13}$
		 &  1.80$\times10^{-11}$ & 3.38$\times10^{-11}$
		  & 2.86$\times10^{4}$ & 6.38$\times10^{3}$
		   \\ 
		~ & 10 & 1.17$\times10^{-3}$ & 4.09$\times10^{-4}$
		 &  1.47$\times10^{-3}$ & 5.01$\times10^{-4}$
		  & 3.26$\times10^{4}$ & 6.55$\times10^{3}$
		   \\ 
		~ & 30 & 3.43$\times10^{1}$ & 1.24$\times10^{1}$
		 & 3.46$\times10^{1}$ & 1.25$\times10^{1}$
		  &  6.11$\times10^{4}$ & 1.12$\times10^{4}$
		  	\\ \hline
		$^{75}$Ge & 3 & 1.56$\times10^{-10}$ & 1.39$\times10^{-10}$
		 &  8.05$\times10^{-9}$ & 7.23$\times10^{-9}$
		  & 5.21$\times10^{4}$ & 8.71$\times10^{3}$
		  \\ 
		~ & 10 & 1.91$\times10^{-3}$ & 1.98$\times10^{-3}$
		 & 2.39$\times10^{-3}$ & 2.47$\times10^{-3}$
		  & 7.24$\times10^{4}$ & 9.02$\times10^{3}$
		   \\ 
		~ & 30 & 1.60$\times10^{2}$ & 2.63$\times10^{1}$
		 &  1.61$\times10^{2}$ & 2.65$\times10^{1}$
		  & 2.98$\times10^{5}$ & 1.77$\times10^{4}$
		  \\ \hline
		$^{76}$Ge & 3 & 1.69$\times10^{-15}$ & 2.90$\times10^{-15}$
		 & 8.77$\times10^{-14}$ & 1.51$\times10^{-13}$
		  &  3.00$\times10^{4}$ & 5.36$\times10^{3}$
		  \\ 
		~ & 10 & 5.14$\times10^{-4}$ & 1.07$\times10^{-4}$
		 &  6.46$\times10^{-4}$ & 1.33$\times10^{-4}$
		  & 3.49$\times10^{4}$ & 5.38$\times10^{3}$
		   \\ 
		~ & 30 & 3.39$\times10^{1}$ & 8.46$\times10^{0}$
		 &  3.42$\times10^{1}$ & 8.52$\times10^{0}$
		  & 6.40$\times10^{4}$ & 9.08$\times10^{3}$
		  \\ \hline
	$^{77}$Ge & 3 & 2.65$\times10^{-13}$ & 1.05$\times10^{-13}$ &
	 1.38$\times10^{-11}$ & 5.46$\times10^{-12}$
	  & 4.51$\times10^{4}$ & 6.32$\times10^{3}$
	   \\ 
	~ & 10 & 8.95$\times10^{-4}$ & 2.99$\times10^{-4}$
	
	 & 1.12$\times10^{-3}$ & 3.74$\times10^{-4}$
	 
	 & 7.41$\times10^{4}$ & 6.56$\times10^{3}$
	  \\ 
	~ & 30 & 1.75$\times10^{2}$ & 1.52$\times10^{1}$
	&  1.77$\times10^{2}$ & 1.53$\times10^{1}$
	 & 3.37$\times10^{5}$ & 1.23$\times10^{4}$
	 \\ \hline
		$^{78}$Ge & 3 & 1.01$\times10^{-16}$ & 1.82$\times10^{-17}$
		 &  5.26$\times10^{-15}$ & 9.46$\times10^{-16}$
		  & 3.33$\times10^{4}$ & 3.75$\times10^{3}$
		  \\ 
		~ & 10 & 4.22$\times10^{-4}$ & 2.22$\times10^{-5}$
		&  5.30$\times10^{-4}$ & 2.76$\times10^{-5}$
		 & 3.79$\times10^{4}$ & 3.94$\times10^{3}$
		  \\ 
		~ & 30 & 3.37$\times10^{1}$ & 5.99$\times10^{0}$
		 & 3.40$\times10^{1}$ & 6.03$\times10^{0}$
		  & 6.40$\times10^{4}$ & 7.00$\times10^{3}$
		  \\ \hline
		$^{79}$Ge & 3 & 5.93$\times10^{-16}$ & 4.17$\times10^{-16}$
		& 3.08$\times10^{-14}$ & 2.16$\times10^{-14}$
		 & 4.32$\times10^{4}$ & 4.79$\times10^{3}$
		 \\ 
		~ & 10 & 8.55$\times10^{-4}$ & 6.25$\times10^{-5}$
		& 1.07$\times10^{-3}$ & 7.83$\times10^{-5}$
		 & 8.04$\times10^{4}$ & 5.01$\times10^{3}$
		  \\ 
		~ & 30 & 1.17$\times10^{2}$ & 9.69$\times10^{0}$
		& 1.78$\times10^{2}$ & 9.76$\times10^{0}$
		& 3.37$\times10^{5}$ & 9.25$\times10^{3}$
		\\ \hline
		$^{80}$Ge & 3 & 7.13$\times10^{-21}$ & 1.51$\times10^{-21}$ &
		 3.71$\times10^{-19}$ & 7.83$\times10^{-20}$ &
		  4.37$\times10^{2}$ & 2.36$\times10^{3}$
		  \\ 
		~ & 10 & 1.19$\times10^{-4}$ & 2.64$\times10^{-6}$
		 & 1.50$\times10^{-4}$ & 3.22$\times10^{-6}$
		  & 1.11$\times10^{4}$ & 2.50$\times10^{3}$
		  \\ 
		~ & 30 &  3.01$\times10^{1}$ & 3.12$\times10^{0}$
		& 3.03$\times10^{1}$ & 3.15$\times10^{0}$
		 & 5.68$\times10^{4}$ & 4.63$\times10^{3}$
		 \\ \hline		
	\end{tabular}
\end{threeparttable}

\newpage
\begin{threeparttable}[H]
	\tiny
	\centering
	\caption{Same as Table \ref{Tab 6} but for ($\beta^{-}$+$PC$) rates.} \label{Tab 7}
	\addtolength{\tabcolsep}{1pt}
\begin{tabular}{|c|c|cc |cc |cc |}
	\hline
	\textbf{Nuclei} & \textbf{$T_9$} &\multicolumn{2}{c|}{ \textbf{$\rho Y_{e}=10^3$}} &\multicolumn{2}{c|}{ \textbf{$\rho Y_{e}=10^7$}}  & \multicolumn{2}{c|}{\textbf{$\rho Y_{e}=10^{11}$}}  \\  \hline
	~ & & $\lambda_{(\beta^{-}+PC)}^{pn-QRPA}$& $\lambda_{(\beta^{-}+PC)} ^{PF03}$ & $\lambda_{(\beta^{-}+PC)}^{pn-QRPA}$ & $\lambda_{(\beta^{-}+PC)} ^{PF03}$ & $\lambda_{(\beta^{-}+PC)}^{pn-QRPA}$ & $\lambda_{(\beta^{-}+PC)} ^{PF03}$ \\ \hline
		$^{67}$Ge & 3 & 7.62$\times10^{-14}$ & 1.24$\times10^{-12}$
		&  1.47$\times10^{-15}$ & 2.48$\times10^{-14}$
		& 4.89$\times10^{-54}$ & 8.91$\times10^{-53}$
		\\ 
		~ & 10 & 1.19$\times10^{-4}$ & 1.00$\times10^{-4}$
		& 9.46$\times10^{-5}$ & 8.04$\times10^{-5}$
		 & 1.16$\times10^{-16}$ & 1.04$\times10^{-16}$
		  \\ 
		~ & 30 & 1.41$\times10^{1}$ & 5.98$\times10^{0}$
		& 1.40$\times10^{1}$ & 6.00$\times10^{0}$
		 & 1.93$\times10^{-3}$ & 8.36$\times10^{-4}$
		 \\ \hline		
		$^{68}$Ge & 3 & 1.87$\times10^{-16}$ & 8.04$\times10^{-15}$
		&  3.61$\times10^{-18}$ & 1.63$\times10^{-16}$
		 & 1.20$\times10^{-56}$ & 1.00$\times10^{-54}$
		 \\ 
		~ & 10 & 9.74$\times10^{-5}$ & 3.43$\times10^{-4}$
		&  7.76$\times10^{-5}$ & 2.74$\times10^{-4}$
		 & 9.50$\times10^{-17}$ & 3.59$\times10^{-16}$
		  \\ 
		~ & 30 & 5.83$\times10^{0}$ & 5.01$\times10^{0}$
		 & 5.79$\times10^{0}$ & 4.97$\times10^{0}$
		  & 7.96$\times10^{-4}$ & 6.93$\times10^{-4}$
		  \\ \hline
		$^{69}$Ge & 3 & 8.36$\times10^{-11}$ & 2.35$\times10^{-9}$ &
		 1.62$\times10^{-12}$ & 4.82$\times10^{-11}$
		  & 5.43$\times10^{-51}$ & 1.68$\times10^{-49}$
		  \\ 
		~ & 10 & 5.91$\times10^{-4}$ & 1.03$\times10^{-3}$
		 &	4.70$\times10^{-4}$ & 8.23$\times10^{-4}$
		  & 5.76$\times10^{-6}$ & 1.08$\times10^{-15}$
		  \\ 
		~ & 30 & 2.42$\times10^{1}$ & 9.27$\times10^{0}$
		 &	2.40$\times10^{1}$ & 9.18$\times10^{0}$
		  & 3.30$\times10^{-3}$ & 1.29$\times10^{-3}$
		  \\ \hline
		$^{70}$Ge & 3 & 6.07$\times10^{-14}$ & 1.51$\times10^{-11}$
		& 1.17$\times10^{-15}$ & 3.02$\times10^{-13}$
		& 3.89$\times10^{-54}$ & 1.08$\times10^{-51}$
		\\ 
		~ & 10 & 3.29$\times10^{-4}$ & 2.71$\times10^{-3}$
		 &  2.62$\times10^{-4}$ & 2.19$\times10^{-3}$
		  & 3.21$\times10^{-16}$ & 2.84$\times10^{-15}$
		  \\ 
		~ & 30 & 8.24$\times10^{0}$ & 7.59$\times10^{0}$
		 & 8.18$\times10^{0}$ & 7.52$\times10^{0}$
		  & 1.12$\times10^{-3}$ & 1.05$\times10^{-3}$
		  \\ \hline
		$^{71}$Ge & 3  & 3.40$\times10^{-8}$ & 2.10$\times10^{-6}$
		&	6.81$\times10^{-10}$ & 4.26$\times10^{-8}$
		 & 5.79$\times10^{-48}$ & 1.51$\times10^{-46}$
		 \\ 
		~ & 10 &  3.12$\times10^{-3}$ & 6.31$\times10^{-3}$
		&	2.49$\times10^{-3}$ & 5.05$\times10^{-3}$
		 &  3.08$\times10^{-17}$ & 6.58$\times10^{-15}$
		  \\ 
		~ & 30 & 6.10$\times10^{1}$ & 1.25$\times10^{1}$
		 &	6.05$\times10^{1}$ & 1.24$\times10^{1}$
		  & 8.34$\times10^{-3}$ & 1.74$\times10^{-3}$
		   \\ \hline
		$^{72}$Ge & 3 & 2.52$\times10^{-11}$ & 1.85$\times10^{-8}$
		 & 5.10$\times10^{-13}$ & 3.66$\times10^{-10}$
		  &  1.72$\times10^{-51}$ & 1.34$\times10^{-48}$
		  
		\\
		~ & 10 & 1.36$\times10^{-3}$ & 1.41$\times10^{-2}$
		 &  1.08$\times10^{-3}$ & 1.13$\times10^{-2}$
		  & 1.34$\times10^{-15}$ & 1.46$\times10^{-14}$
		  
		\\ 
		~ & 30 & 1.30$\times10^{1}$ & 1.01$\times10^{1}$
		 &	1.29$\times10^{1}$ & 1.00$\times10^{1}$
		  & 1.77$\times10^{-3}$ & 1.40$\times10^{-3}$
		   \\ \hline
		$^{73}$Ge & 3 & 4.66$\times10^{-6}$ & 7.21$\times10^{-6}$
		 & 3.45$\times10^{-7}$ & 1.73$\times10^{-6}$
		  & 3.77$\times10^{-44}$ & 2.68$\times10^{-46}$
		   \\ 
		~ & 10 & 8.32$\times10^{-3}$ & 1.25$\times10^{-2}$
		 & 	 6.74$\times10^{-3}$ & 1.02$\times10^{-2}$
		  & 1.63$\times10^{-14}$  & 2.25$\times10^{-14}$
		   \\ 
		~ & 30 & 7.11$\times10^{1}$ & 1.50$\times10^{1}$
		 &	7.06$\times10^{1}$ & 1.49$\times10^{1}$
		 & 9.73$\times10^{-3}$ & 2.09$\times10^{-3}$
		  \\ \hline
		$^{74}$Ge & 3 & 6.25$\times10^{-9}$ & 2.74$\times10^{-6}$
		 &  5.31$\times10^{-10}$ & 5.59$\times10^{-8}$
		  & 8.03$\times10^{-48}$ & 1.96$\times10^{-46}$
		   \\ 
		~ & 10 & 4.76$\times10^{-3}$ & 9.76$\times10^{-2}$
		 & 3.81$\times10^{-3}$ & 7.80$\times10^{-2}$
		  & 4.99$\times10^{-15}$ & 1.01$\times10^{-13}$
		   \\ 
		~ & 30 & 2.46$\times10^{1}$ & 1.46$\times10^{1}$
		 &  2.44$\times10^{1}$ & 1.45$\times10^{1}$
		  & 3.37$\times10^{-3}$ & 2.02$\times10^{-3}$
		  	\\ \hline
		$^{75}$Ge & 3 & 2.69$\times10^{-4}$ & 6.68$\times10^{-4}$
		 & 1.06$\times10^{-4}$ & 3.89$\times10^{-4}$
		  & 6.94$\times10^{-42}$ & 2.27$\times10^{-45}$
		  \\ 
		~ & 10 & 3.15$\times10^{-2}$ & 5.12$\times10^{-2}$
		 & 2.56$\times10^{-2}$ & 4.50$\times10^{-2}$
		  & 6.94$\times10^{-42}$ & 4.20$\times10^{-13}$
		  \\ 
		~ & 30 & 1.73$\times10^{2}$ & 2.02$\times10^{1}$
		 & 1.71$\times10^{2}$ & 2.00$\times10^{1}$
		  & 2.37$\times10^{-2}$ & 2.85$\times10^{-3}$
		  \\ \hline
		$^{76}$Ge & 3 &  7.37$\times10^{-7}$ & 5.04$\times10^{-5}$
		 &  6.73$\times10^{-8}$ & 1.45$\times10^{-5}$
		  & 4.39$\times10^{-44}$ & 9.91$\times10^{-46}$
		  \\ 
		~ & 10 & 1.78$\times10^{-2}$ & 2.73$\times10^{-1}$
		 &  1.47$\times10^{-2}$ & 2.22$\times10^{-1}$
		  &  5.10$\times10^{-14}$ & 3.74$\times10^{-13}$
		  \\ 
		~ & 30 & 3.89$\times10^{1}$ & 1.91$\times10^{1}$
		 &  3.86$\times10^{1}$ & 1.89$\times10^{1}$
		  & 5.34$\times10^{-3}$ & 2.65$\times10^{-3}$
		  \\ \hline
		$^{77}$Ge & 3 &  3.46$\times10^{-3}$ & 9.61$\times10^{-3}$
		 &  2.63$\times10^{-3}$ & 8.09$\times10^{-3}$
		  & 6.21$\times10^{-39}$ & 2.97$\times10^{-45}$
		  \\ 
		~ & 10 & 9.62$\times10^{-2}$ & 2.44$\times10^{-1}$
		 &  8.13$\times10^{-2}$ & 2.32$\times10^{-1}$
		  & 1.04$\times10^{-12}$ & 8.40$\times10^{-12}$
		    \\ 
		~ & 30 & 2.86$\times10^{2}$ & 2.47$\times10^{1}$
		&  2.83$\times10^{2}$ & 2.46$\times10^{1}$
		 & 3.94$\times10^{-2}$ & 3.70$\times10^{-3}$
		 \\ \hline
		$^{78}$Ge & 3 &  3.67$\times10^{-5}$ & 1.18$\times10^{-3}$
		 &  1.53$\times10^{-6}$ & 8.23$\times10^{-4}$
		  & 5.32$\times10^{-41}$ & 3.20$\times10^{-45}$
		  \\ 
		~ & 10 & 6.69$\times10^{-2}$ & 7.00$\times10^{-1}$
		 &  5.69$\times10^{-2}$ & 6.28$\times10^{-1}$
		  &  8.18$\times10^{-13}$ & 6.05$\times10^{-12}$
		   \\ 
		~ & 30 & 6.15$\times10^{1}$ & 2.39$\times10^{1}$
		 &  6.10$\times10^{1}$ & 2.37$\times10^{1}$
		  & 8.50$\times10^{-3}$ & 3.35$\times10^{-3}$
		  \\ \hline
		$^{79}$Ge & 3 & 5.31$\times10^{-3}$ & 8.72$\times10^{-2}$
		 &  4.46$\times10^{-3}$ & 8.00$\times10^{-2}$
		  &  2.05$\times10^{-36}$ & 5.66$\times10^{-44}$
		  \\ 
		~ & 10 & 2.09$\times10^{-1}$ & 1.03$\times10^{0}$
		 &  1.86$\times10^{-1}$ & 1.00$\times10^{0}$
		  &  8.66$\times10^{-12}$ & 1.09$\times10^{-10}$
		   \\ 
		~ & 30 & 2.85$\times10^{2}$ & 3.21$\times10^{1}$
		 &  2.83$\times10^{2}$ & 3.19$\times10^{1}$
		  & 3.95$\times10^{-2}$ & 5.61$\times10^{-3}$
		  \\ \hline
		$^{80}$Ge & 3 & 7.23$\times10^{-4}$ & 4.05$\times10^{-2}$
		 &  2.73$\times10^{-4}$ & 3.45$\times10^{-2}$
		  & 1.40$\times10^{-38}$ & 1.23$\times10^{-44}$
		  \\ 
		~ & 10 & 1.35$\times10^{-1}$ & 3.90$\times10^{0}$
		 &  1.17$\times10^{-1}$ & 3.73$\times10^{0}$
		  & 4.92$\times10^{-12}$ & 1.50$\times10^{-10}$
		   \\ 
		~ & 30 & 7.52$\times10^{1}$ & 3.02$\times10^{1}$
		&  7.47$\times10^{1}$ & 2.99$\times10^{1}$
		 & 1.05$\times10^{-2}$ & 4.46$\times10^{-3}$
		 \\ \hline		
	\end{tabular}
\end{threeparttable}

 \end{document}